\documentclass[fleqn,usenatbib]{mnras}

\usepackage{newtxtext,newtxmath}
\usepackage[T1]{fontenc}

\DeclareRobustCommand{\VAN}[3]{#2}
\let\VANthebibliography\thebibliography
\def\thebibliography{\DeclareRobustCommand{\VAN}[3]{##3}\VANthebibliography}

\usepackage{graphicx}	% Including figure files
\usepackage{amsmath}	% Advanced maths commands
\title[Hot gas galaxy halo]{Exploring the hot gaseous halo around an extremely massive and relativistic jet launching spiral galaxy with \textit{XMM$-$Newton}}
\author[M. S. Mirakhor et al.]{
\parbox[]{6.5in} {M. S. Mirakhor,$^{1}$\thanks{Email: msm0033@uah.edu}
S. A. Walker,$^{1}$ 
J. Bagchi,$^{2}$
A. C. Fabian,$^{3}$
A. J. Barth,$^{4}$
F. Combes,$^{5,6}$
P. Dabhade,$^{7,2}$
L. C. Ho,$^{8,9}$
and M. B. Pandge$^{10}$}
\\
$^{1}$Department of Physics and Astronomy, The University of Alabama in Huntsville, Huntsville, AL 35899, USA \\
$^{2}$The Inter-University Centre for Astronomy and Astrophysics (IUCAA), Pune University Campus, Post Bag 4, Pune 411007, India\\
$^{3}$Institute of Astronomy, Madingley Road, Cambridge CB3 0HA, UK\\
$^{4}$Department of Physics and Astronomy, University of California, Irvine, 4129 Frederick Reines Hall, Irvine, CA 92697, USA\\
$^{5}$Observatoire de Paris, LERMA, CNRS, PSL Universit\'e, Sorbonne Universit\'e, F-75014 Paris, France\\
$^{6}$Coll\`ege de France, 11 Place Marcelin Berthelot, F-75005 Paris, France\\
$^{7}$Leiden Observatory, Leiden University, P.O. Box 9513, NL-2300 RA, Leiden, The Netherlands\\
$^{8}$Kavli Institute for Astronomy and Astrophysics, Peking University, Beijing 100871, People’s Republic of China\\
$^{9}$Department of Astronomy, School of Physics, Peking University, Beijing 100871, People’s Republic of China\\
$^{10}$Dayanand Science College, Barshi Road, Latur, Maharashtra 413512, India
}

\date{Accepted XXX. Received YYY; in original form ZZZ}

\pubyear{2020}

\begin{document}
\label{firstpage}
\pagerange{\pageref{firstpage}--\pageref{lastpage}}
\maketitle

\begin{abstract}
We present a deep \textit{XMM$-$Newton} observation of the extremely massive, rapidly rotating, relativistic-jet-launching spiral galaxy 2MASX J23453268-0449256. Diffuse X-ray emission from the hot gaseous halo around the galaxy is robustly detected out to a radius of 160 kpc, corresponding roughly to 35 per cent of the virial radius ($\approx 450$ kpc). We fit the X-ray emission with the standard isothermal $\beta$ model, and it is found that the enclosed gas mass within 160 kpc is $1.15_{-0.24}^{+0.22} \times 10^{11} \, \rm{M}_{\odot}$. Extrapolating the gas mass profile out to the virial radius, the estimated gas mass is $8.25_{-1.77}^{+1.62} \times 10^{11} \, \rm{M}_{\odot}$, which makes up roughly 65 per cent of the total baryon mass content of the galaxy. When the stellar mass is considered and accounting for the statistical and systematic uncertainties, the baryon mass fraction within the virial radius is $0.121_{-0.043}^{+0.043}$, in agreement with the universal baryon fraction. The baryon mass fraction is consistent with all baryons falling within $r_{200}$, or with only half of the baryons falling within $r_{200}$. Similar to the massive spiral galaxies NGC 1961 and NGC 6753, we find a low value for the metal abundance of $\approx 0.1 {\rm{Z}}_{\odot}$, which appears uniform with radius. We also detect diffuse X-ray emission associated with the northern and southern lobes, possibly attributed to inverse Compton scattering of cosmic microwave background photons. The estimated energy densities of the electrons and magnetic field in these radio lobes suggest that they are electron-dominated by a factor of 10$-$200, depending on the choice of the lower cut-off energy of the electron spectrum.    

\end{abstract}


%%%%%%%%%%%%%%%%%%%%%%%%%%%%%%%%%%%%%%%%%%%%%%%%%%

%%%%%%%%%%%%%%%%% BODY OF PAPER %%%%%%%%%%%%%%%%%%

\section{Introduction}
Since \citet{White1978}, models of galaxy formation have predicted that galaxies should be surrounded by hot gaseous halos. These hot halos are formed as matter accretes onto the dark matter halo, with shocks expected to heat up the baryons to the virial temperature \citep{White1991,Benson2010}. The hot halos are predicted to be a significant source of baryons, containing as much or even more baryonic mass than the galaxies within the halos \citep{Sommer-Larsen2006,Fukugita2006}.

When compared to the mean cosmic baryon to matter ratio determined by Planck \citep[$0.156 \pm 0.003$;][]{ade2016planck}, observations indicate that nearby galaxies are missing most of their baryons \citep[e.g.][]{Hoekstra2005,Heymans2006,Bregman2007}. This lack of baryons has been confirmed in other galaxies \citep[e.g.][]{McGaugh2005,mcgaugh2009baryon} using a variety of methods. However, the baryon mass in the hot gas halos of galaxies has not been included in most of these studies. It is possible that the majority of the missing baryons in galaxies could reside in these hot halos.

Hot halos around early-type galaxies have been well studied in soft X-rays \citep[e.g.][]{Forman1985,OSullivan2001}. However, the histories of these halos are complex, as coronal gas can also be produced in the mergers and star formation occurring as the galaxy became an elliptical \citep{Read1998}.  It is also difficult to distinguish the halo gas from the intergroup medium in which most large ellipticals are located \citep{Dressler1980}.

One exciting avenue for study is massive disk galaxies, which have not undergone major merger events, and which are sufficiently massive that their hot halo is bright enough in X-rays to be studied with \textit{Chandra} and \textit{XMM$-$Newton}. Recent breakthroughs in the observation of hot halos around disk galaxies have been made using \textit{Chandra} and \textit{XMM$-$Newton} to detect the hot halos around the extremely massive, fast-rotating spiral galaxies NGC 1961 (\citealt{Anderson2011}; \citealt{Bogdan2013a}; \citealt{anderson2016deep}), UGC 12591 \citep{Dai2012}, NGC 266 (\citealt{Bogdan2013b}), NGC 6753 (\citealt{Bogdan2013a}; \citealt{bogdan2017}), and 2MASX J23453268-0449256 (\citealt{Walker2015a}). 

The object 2MASX J23453268-0449256 (hereafter J2345-0449) is an extremely massive, rapidly rotating, relativistic-jet-launching spiral galaxy. Its radio and optical properties have been studied by \citet{Bagchi2014}, using the IUCAA Girawali Observatory 2m telescope. The kinematics of the optical Balmer H$\alpha$ line reveals an extremely large rotation speed, $V_{\rm{rot}}=371/\sin(i)=429 \pm 30$ km s$^{-1}$, in the asymptotic flat region at $r \geq 10$ kpc from the galactic center. J2345-0449 is therefore one of the most massive known spiral galaxies. \citet{Bagchi2014} also found that the central region of the galaxy (3 arcsec, corresponding to 4.3 kpc) has an exceptionally large stellar velocity dispersion, $\sigma=326 \pm 59$ km s$^{-1}$. On such a spatial scale, this is higher than for the majority of bulge-less disks. This implies a huge concentration of mass of $10^{11}\,\rm{M}_{\odot}$ within the central region, including a supermassive black hole (SMBH), for which the lower-limit mass from the optical data is $2 \times 10^8\,\rm{M}_{\odot}$.

This galaxy is currently ejecting a collimated pair of relativistic jets out to large radii \citep{Bagchi2014}. Furthermore, \citet{Bagchi2014} detected the synchrotron radio emission that arises from megaparsec-scale bipolar structure with two nearly aligned pairs of radio lobes. The inner and outer radio lobe pairs are remarkably large, and their emission extends over $\approx 1.6$ Mpc. These radio lobes, however, lack prominent hot spots and are no longer being energized by the jets. 

A close alignment has been found between the inner and outer radio lobe pairs, implying a stable spin axis of the black hole over the timescale of $\approx 10^8$ yr between the two-last episodes of jet triggering. \citet{Bagchi2014} also found that this spiral galaxy features a pseudo-bulge rather than a classical bulge. These features lead \citet{Bagchi2014} to suggest that the galactic disk and its SMBH have evolved together quietly, and have not undergone recent major merger events. This suggestion is also driven by the absence of any tidal debris (such as tails, shells, or plumes), the location of the galaxy in an isolated galactic environment with no nearby galactic neighbors, and the fact that the galaxy has highly-symmetric spiral arms within a rotationally supported disk.

Using a 100 ks \textit{Chandra} observation, \citet{Walker2015a} detected extended X-ray emission from the hot gaseous halo surrounding the spiral galaxy. \citet{Walker2015a} found that this X-ray emission is elongated along the plane of the galaxy disc, and extends out to a radius of $\approx 80$ kpc, far beyond the galaxy's optical radius of $\approx 25$ kpc. Furthermore, the \textit{Chandra} data also revealed extended X-ray features, coinciding with the inner and outer radio lobes, and it is possibly due to inverse Compton (IC) scattering of cosmic microwave background (CMB) photons. Under the assumption of spherical symmetry and fitting the emission with the standard isothermal $\beta$ model, \citet{Walker2015a} estimated the hot halo mass of $2.0_{-1.0}^{+1.0} \times 10^{10} \, \rm{M}_{\odot}$ for this galaxy within 80 kpc, the maximum radius out to which the X-ray emission is detected by \textit{Chandra}.

However, the low effective area of \textit{Chandra} means that the number of counts collected from the hot halo of the galaxy is low (around 130 counts), which is too low to determine the hot halo temperature and metal abundance, both of which are required to accurately measure the hot gas mass in the halo. In addition, the small field of view of the \textit{Chandra} ACIS-S instrument means that the outer lobes have not been properly explored in X-rays yet. \textit{XMM$-$Newton}, with its 9 times greater collecting area in the soft X-ray band (0.4$-$2.0 keV) and much larger field of view, is the only X-ray telescope capable of observing the entire radio galaxy lobe system and providing sufficient counts for a spectroscopic analysis of the hot halo. 

In this work, we present a deep \textit{XMM$-$Newton} observation of the spiral galaxy J2345-0449 to study its hot gaseous halo, investigate its mass distribution, and map its morphology. In addition, we study the magnetic field and energetics of the northern and southern radio lobes, using X-ray and radio data.

Throughout this paper, we adopt a $\Lambda$ CDM cosmology with $\Omega_{\rm{m}}=0.3$, $\Omega_{\rm{\Lambda}}=0.7$, and $H_0=100\,h_{100}$ km s$^{-1}$ Mpc$^{-1}$ with $h_{100}=0.7$. At the redshift of the galaxy ($z=0.0755$), 1 arcmin corresponds to 85.9 kpc. All uncertainties unless otherwise stated are at the 1$\sigma$ level. 

%----------------------
\begin{figure*}
	\includegraphics[width=0.8\textwidth]{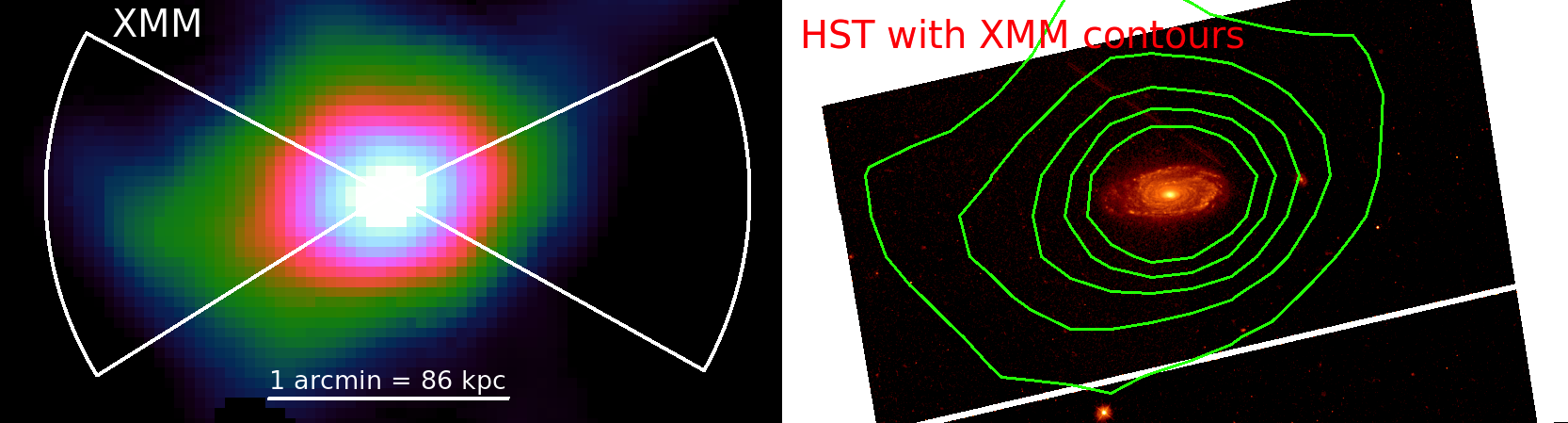}
	\caption{Comparing the background-subtracted, exposure-corrected \textit{XMM$-$Newton} image of the galaxy J2345-0449 in the 0.5$-$1.2 keV band (left-hand panel) to the \textit{HST} image taken with the F438W filter and the UVIS detector (right-hand panel). The X-ray contours are plotted on top of the \textit{HST} image to show the spatial extent of the the X-ray halo. The white sectors in the left-hand panel show the regions that X-ray counts are extracted, avoiding the regions to the north and south of the galaxy where the jets are existed. North is up and east is to the left.}
	\label{fig: xmm_hst_image}
\end{figure*}
%----------------------

\section{Observations and Data Reduction}
\label{sec: data}
\subsection{X-ray data}
The spiral galaxy J2345-0449 was observed with the European Photon Imaging Camera (EPIC) abroad \textit{XMM$-$Newton} for 100 ks between 2017 December 19 and 2017 December 20 (PI: S. A. Walker). The data were reduced using the \textit{XMM$-$Newton} Science Analysis System (XMM-SAS) package version 18.0 and Current Calibration Files (CCF), following the methods outlined in \citet{Snowden2008}. To perform basic data reduction, we run the \textit{epchain} and \textit{emchain} scripts, followed by the \textit{mos-filter} and \textit{pn-filter} tasks to remove soft proton flares and extract calibrated event files from the observations. The individual CCDs in the MOS detectors were then screened and any CCDs that operate in anomalous states were excluded from further processing. We also detected and removed point sources and any extended substructures that contaminated the field of view by running the \textit{cheese} task. We then created the required spectra and response files for the interested region by running \textit{mos-spectra} and \textit{pn-spectra}, and these files were used to create the quiescent particle background spectra and images by running the \textit{mos-back} and \textit{pn-back} tasks. 

Furthermore, we modelled residual soft proton contamination that may have remained after the initial light curve screening using the task \textit{proton}. We also carried out an additional filtering step by running the \textit{Chandra}'s source-detection tool \textit{wavdetect} to detect remaining point sources within the field that were missed using the \textit{cheese} tool. For this purpose, we used an exposure-corrected image in the hard energy band of the galaxy’s selected region that obtained using the \textit{Chandra} data \citep{Walker2015a}. The detected point sources in the \textit{Chandra} image were then excluded from our analysis.

The analysis procedure described above created all required components for a exposure-corrected and background-subtracted image. After weighting each detector by its effective area, these components were then combined and adaptively smoothed into a single image. 

\subsection{Radio data}
J2345-0449 (Proposal code: 29-061, PI: J. Bagchi) was observed with Giant Metre-wave Radio Telescope \citep[GMRT;][]{swarup1991giant} on 2015 November 7 (observation ID: 8151) at 610 MHz with 32 MHz bandwidth. 3C 48 was observed as the primary calibrator for flux calibration at the start and end of the observation. The target (J2345-0449) was observed for a total of 294 min with periodic scans of 25 min each, and source 0022+002 was observed for 5 min as the secondary calibrator between the scans of the target for phase calibration.

The GMRT 610 MHz radio data was analyzed by following the steps of flagging of the RFI calibration, averaging the data and imaging with self-calibration. This was done by using the package Source Peeling and Atmospheric Modelling \citep[SPAM;][]{intema2009ionospheric,intema2017gmrt}, which is based on NRAO’s
Astronomical Image Processing System (AIPS) with a \textsc{python} language interface. This package also provides direction-dependent calibration to correct the effects of the ionosphere, and hence improves the quality of the radio images.

\section{Results}
\label{sec: results}

\subsection{Images}
In the left-hand panel of Fig. \ref{fig: xmm_hst_image}, we show the background-subtracted and exposure-corrected \textit{XMM$-$Newton} image of the spiral galaxy J2345-0449 in the 0.5$-$1.2 keV energy band. This soft band was chosen to maximize the signal-to-noise ratio of the detection of the hot halo surrounding the spiral galaxy. The image is smoothed in such a way that the number of events used for the kernel is 100 counts, and the typical radius of the extracted point sources is in the range of 10-20 arcsec. The right-hand panel of Fig. \ref{fig: xmm_hst_image} shows the \textit{Hubble Space Telescope} (\textit{HST}) image of the galaxy with the \textit{XMM$-$Newton} X-ray contours of the hot halo overplotted. We see that the X-ray emission from the hot halo has much greater spatial extent than the optical emission. The detailed analysis of the galaxy's structure from the \textit{HST} data will be presented in a separate paper (Bagchi et al.; in preparation).

\begin{figure*}
	\includegraphics[width=0.8\textwidth]{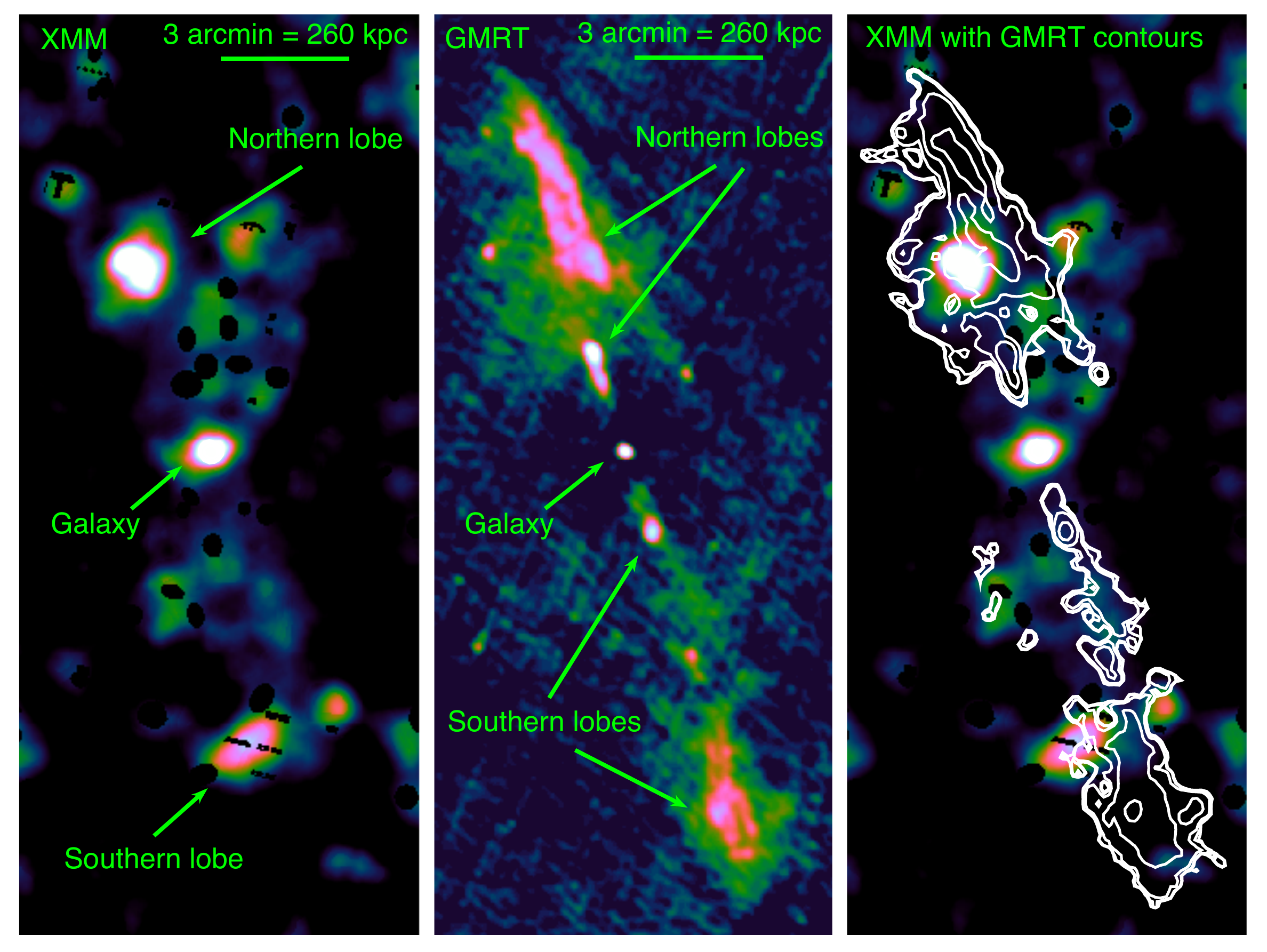}
	\caption{\textit{Left}: Background-subtracted, exposure-corrected X-ray image of the spiral galaxy J2345-0449 in the 0.5$-$1.2 keV energy band. \textit{Middle}: 610 MHz radio image of the galaxy J2345-0449 taken with GMRT. \textit{Right}: X-ray image of J2345-0449 with the 610 MHz GMRT radio contours being overlaid.}
	\label{fig: xmm_gmrt_image}
\end{figure*}

The left-hand panel of Fig. \ref{fig: xmm_gmrt_image} shows the wide-scale \textit{XMM$-$Newton} image of J2345-0449. The image features two extended X-ray structures, positioned at distances of about 400 kpc and 590 kpc, respectively, to the north and south from the galaxy centre. In the middle panel, we show the 610 MHz radio image of the galaxy J2345-0449 taken with GMRT. The extended X-ray emission in the north is coincident with the location of the northern radio lobe, whereas the south X-ray emission is not totally associated with the southern radio lobe (Fig. \ref{fig: xmm_gmrt_image}, right-hand panel), suggesting a more complicated relationship between the X-ray and radio lobe emissions.

%----------------------
%----------------------

\subsection{Surface brightness and density}
\label{sec: surface_brightness}
\begin{figure}
	\includegraphics[width=\columnwidth]{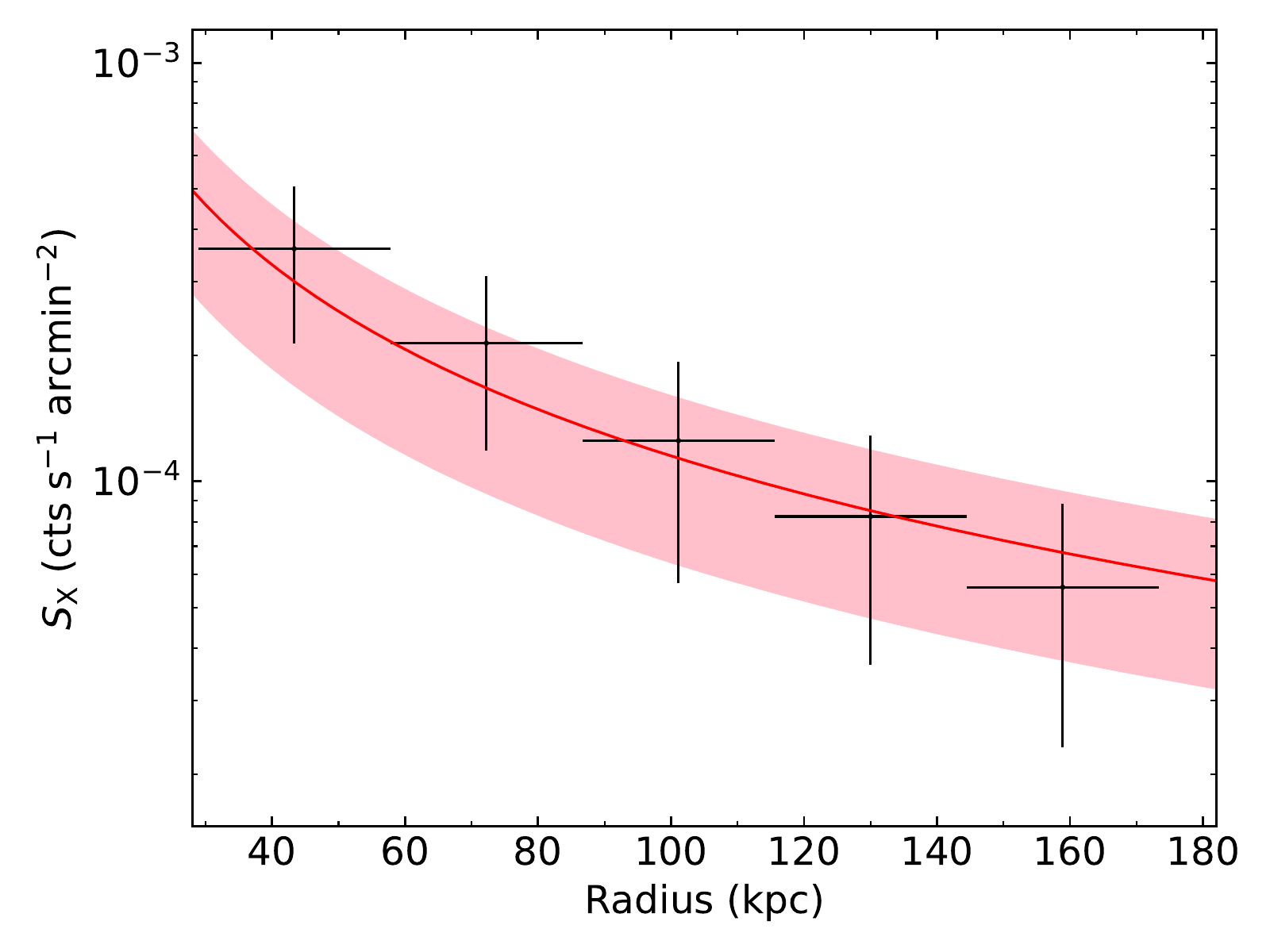}
	\caption{Background-subtracted surface brightness profile of the J2345-0449 galaxy in the 0.5$-$1.2 keV energy band. The red solid line is the best-fitting model, and the shadow pink region marks the $1\sigma$ error computed using a Monte Carlo technique.}
	\label{fig: surface_brightness}
\end{figure}

To derive surface brightness profile of the spiral galaxy J2345-0449, counts were extracted in concentric annuli centred at the cluster centre, (RA, Dec.) = (23:45:32.60, $-$04:49:25.87), corresponding to the position of the peak X-ray flux. To maximize the signal-to-noise ratio of the surface brightness profile of the hot halo, we extracted counts in the soft band (0.5$-$1.2 keV), considering only the directions far away from the jets, shown by the two white sectors in Fig. \ref{fig: xmm_hst_image}. X-ray counts were also extracted from the local background away from the galaxy and jets. In Fig. \ref{fig: surface_brightness}, we show the radial profile of the background-subtracted surface brightness of the galaxy J2345-0449, avoiding the regions to the north and south of the galaxy where the jets are. We find that the measured surface brightness profile robustly traces the gas out to about 160 kpc, out to about 35 per cent of the virial radius.   

%----------------------
%----------------------

Using the \textit{Chandra} data, \citet{Walker2015a} calculated the contribution of the extended X-ray emission from the Low Mass X-ray Binaries (LMXBs) in the same region of J2345-0449 by extracting counts in the 2.0$-$6.0 keV band, in which the emission is expected to be dominated by the LMXB emission. Assuming that the LMXB emission follows a power-law spectrum with an index of 1.56 \citep{Irwin2003}, \citet{Walker2015a} found that the LMXB contamination is negligible in the soft band.   

To parameterise the hot gas distribution of J2345-0449 in the soft band, we fitted the radial profile of the background-subtracted surface brightness with the standard isothermal $\beta$ model \citep{cavaliere1976}: 
%----------------------
\begin{equation}
    S(r)=S_0\bigg[1+\bigg(\frac{r}{r_0}\bigg)^2\bigg]^{0.5-3\beta},
    \label{eq: Sur_bri_beta}
\end{equation}
%----------------------
where $S_0$ is the central surface brightness, $r_0$ is the core radius, and $\beta$ describes the shape of the gas distribution.

For the fitting processes, we used the affine invariant Markov Chain Monte Carlo ensemble sampler implemented in the \textit{emcee} package, as proposed by  \citet{goodman2010ensemble}. Since \textit {Chandra} has a better spatial resolution than \textit{XMM$-$Newton} ($\sim$ 0.5 arcsec for \textit {Chandra} compared to 6 arcsec for \textit{XMM$-$Newton}), we fixed the central parameters at their best values obtained from the fits of the standard $\beta$ model with the \textit{Chandra} data \citep{Walker2015a}, and only allow $\beta$ free to vary. The best-fitting value of the shape parameter, $\beta$, obtained from this fit is $0.36 \pm 0.01$. The predicted value of $\beta$ is consistent well with that reported by \citet{Walker2015a}, with an uncertainty much smaller than that presented by \citet{Walker2015a}. The best-fitting model is shown as the red solid line in Fig. \ref{fig: surface_brightness}. The $1\sigma$ error computed using a Monte Carlo technique, and shown as the shadow pink area in Fig. \ref{fig: surface_brightness}.

Under the assumptions of an isothermal gas and a spherical symmetry, the gas density is given by:
%----------------------
\begin{equation}
    n(r)=n_0\bigg[1+\bigg(\frac{r}{r_0}\bigg)^2\bigg]^{1.5\beta},
    \label{eq: density_beta}
\end{equation}
%----------------------
where $n_0$ is the central density. 

\begin{figure}
	\includegraphics[width=\columnwidth]{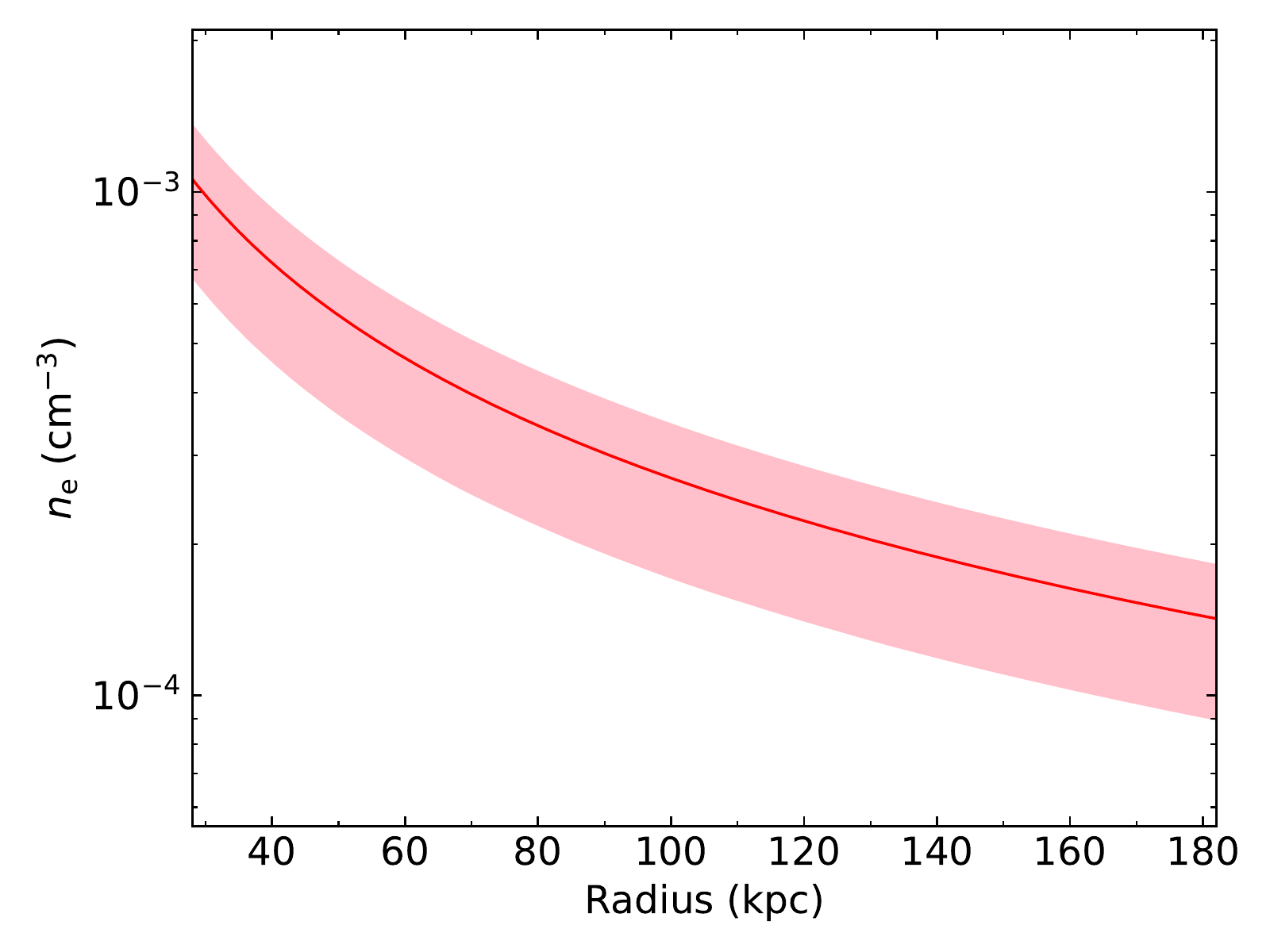}
	\caption{Radial profile of the electron density of the J2345-0449 galaxy obtained using the best-fitting parameters of the $\beta$ model. The shadow pink region marks the $1\sigma$ error computed using a Monte Carlo technique.}
	\label{fig: density_profile}
\end{figure}

Fixing the galaxy redshift to 0.0755 and the absorbing column density to the Leiden–Argentine–Bonn (LAB) survey value of $3\times 10^{20}$ cm$^{-2}$ \citep{kalberla2005leiden}, \citet{Walker2015a} found that the central density $n_0$ is equal to $0.08_{-0.02}^{+0.03}$ cm$^{-3}$. We adopt this value for $n_0$ in this work. In Fig. \ref{fig: density_profile}, we plot the estimated density profile of J2345-0449 using the best-fitting parameters derived from the $\beta$ model.

The X-ray image of J2345-0449 (Fig. \ref{fig: xmm_hst_image}, left-hand panel) shows that the surface brightness is not fully spherical, with the bulk of the X-ray emission extending in the east-west direction along the major axis of the galaxy. To examine the effects of the excluding regions along the jet directions on the shape parameter $\beta$, we fitted the X-ray data along the jet directions with an isothermal $\beta$ model. Within a relatively small radius ($< 80$ kpc), the estimated value of the shape parameter is $\beta \sim 0.36$, consistent well with that derived in the direction of the galaxy plane. However, due to the effects of the active galactic nuclei (AGN) feedback on the hot gas along the jet directions, we therefore included only the X-ray counts in the direction of the plane of the galaxy, far away from the jets.

Furthermore, we quantified the shape of the galaxy and examined its effects on the baryon mass fraction measurements under the assumption of spherical symmetry by fitting the X-ray surface brightness image (Fig. \ref{fig: xmm_hst_image}, left-hand panel) with an isothermal $\beta$ model, and accounting for projected ellipticity $\epsilon$ and the angle of the ellipticity $\theta$. We find a projected ellipticity of $0.31 \pm 0.04$, implying that the ratio of the minor-to-major axis is $\sim 0.7$. Several studies \citep[e.g.][]{mohr1995cosmological,cooray1998cosmology} found that a system with $\epsilon \approx 0.3$ can have an uncertainty of about $\pm 6$ per cent in the gas mass and about $\pm 12$ per cent in the total mass if it is modeled as a sphere. This translates to $\sim \pm 13$ per cent uncertainty in the baryon mass fraction measurements derived from the X-ray data. On the other hand, the isothermal assumption can systematically affect the baryon mass fraction measurements through its effects on both gas mass and total mass. However, since the X-ray gas mass is a weak function of temperature ($\propto T^{1/2}$) and the total mass is approximately proportional to temperature ($\propto T$), consequently, the baryon mass fraction ($\propto T^{-1/2}$) is not very sensitive to temperature gradients in the hot gaseous halo. We therefore assign a $+8$ per cent systematic uncertainty to the baryon mass fraction under the isothermal assumption. Our baryon mass fraction measurements presented in Section \ref{sec: baryon_fraction} take account of the systematic uncertainties due to the effects of asphericity and the isothermal assumption discussed above.        

%----------------------
%----------------------

\subsection{Spectral Analysis}
\label{spectral}

The much higher effective area of \textit{XMM$-$Newton} compared to \textit{Chandra} allows us to perform a spectral analysis of the hot gas in J2345-0449 for the first time. To perform the spectral analysis, we follow the \textit{XMM$-$Newton} Extended Source Analysis (ESAS) cookbook\footnote{https://heasarc.gsfc.nasa.gov/docs/xmm/esas/cookbook/xmm-esas.html}, as is also done in \citet{anderson2016deep}. Spectra were extracted in annuli with width 0.25 arcmin from the MOS and PN detectors using \textit{mos-spectra} and \textit{pn-spectra} respectively, while particle background files were obtained using \textit{mos-back} and \textit{pn-back} respectively. For each annulus, the MOS and PN spectra were fit simultaneously, together with the \textit{ROSAT} All Sky Survey (RASS) spectrum from a background annulus in the region $1.5{-}2.5$ degrees around the galaxy which is used to model the soft background components. These spectra were fit with a model consisting of the following components. Two solar abundance APEC \citep{Smith2001} models were used to model the local bubble and the Galactic halo, whose temperatures were 0.1 and 0.25 keV respectively. An absorbed power law with index 1.44 was used to model the cosmic X-ray background (CXB), where the column density of the absorption model is fixed to the value from the LAB survey (\citealt{kalberla2005leiden}) of $3\times10^{20}$ cm$^{-2}$. We include zero-width Gaussian models with energies fixed at 1.49 keV and 1.75 keV to model the Al K$\alpha$ and the Si K$\alpha$ instrumental lines, with normalizations being free for each detector. Six zero-width Gaussian were also included to model potential solar wind charge exchange emission (SWCX), and the energies of the these lines were fixed to 0.46, 0.57, 0.65, 0.81, 0.92 and 1.35 keV, which correspond to the the C \rm{VI}, O \rm{VII}, O \rm{VIII}, O \rm{VIII}, Ne \rm{IX} and Mg \rm{XI} transitions. The soft proton background is modelled as a power law with free index and normalization for each detector, using a diagonal response matrix. 

\begin{figure*}
\hbox{	\includegraphics[width=\columnwidth]{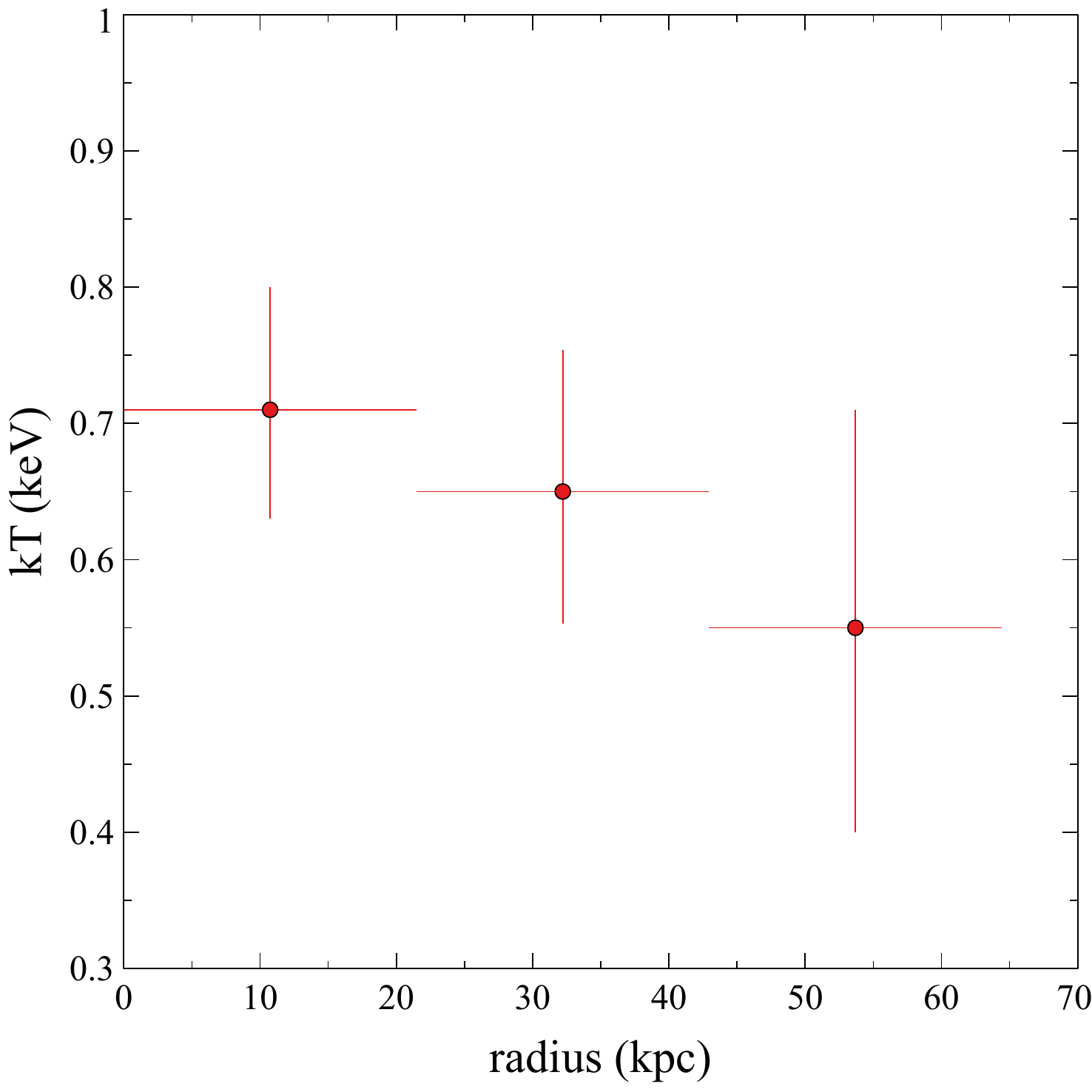}
	\includegraphics[width=\columnwidth]{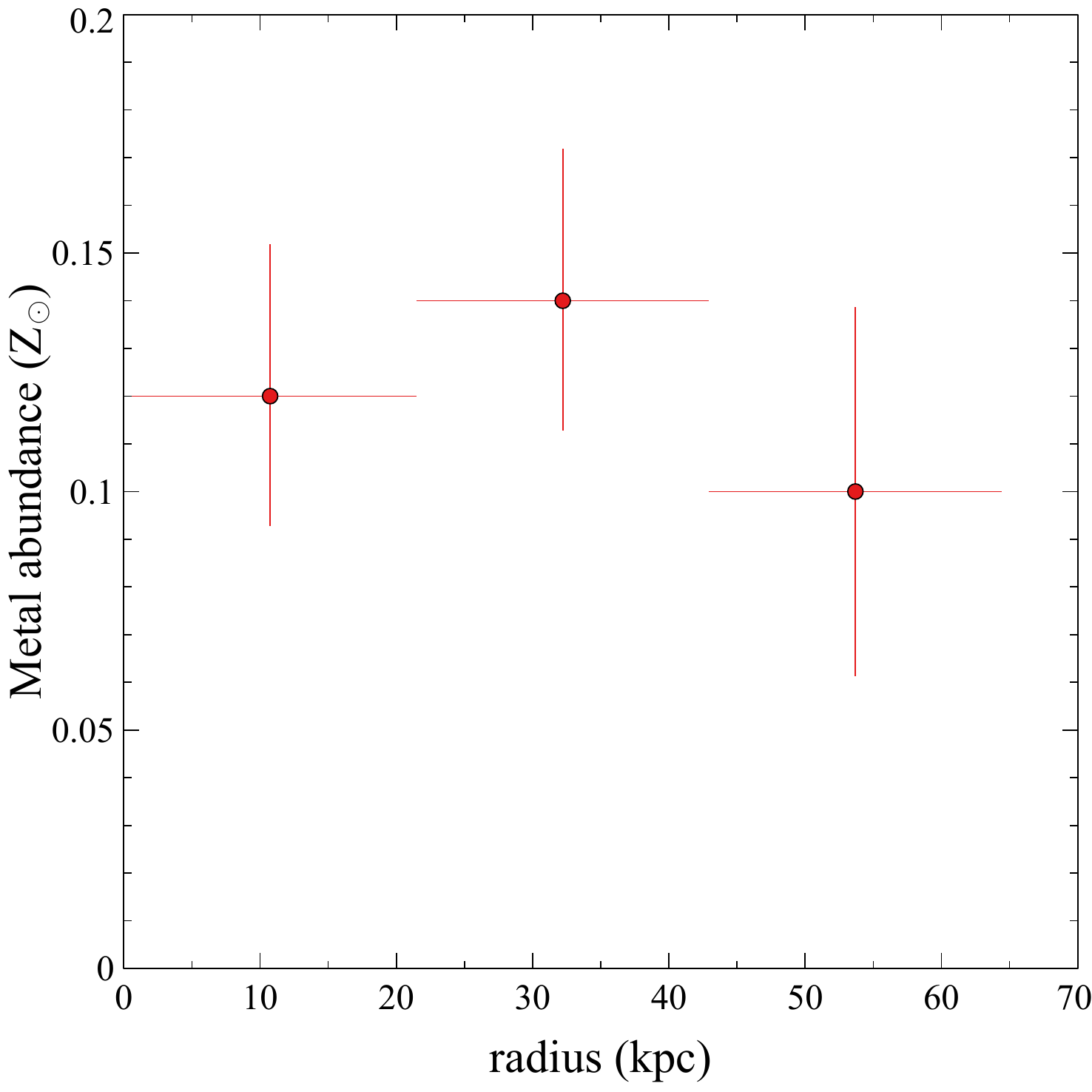}}
	
	\caption{Profiles of the hot halo gas temperature (left) and metal abundance (right) obtained from spectral fitting.}
	\label{fig: temp_profile}
\end{figure*}

\begin{figure}
\vbox{\includegraphics[width=\columnwidth]{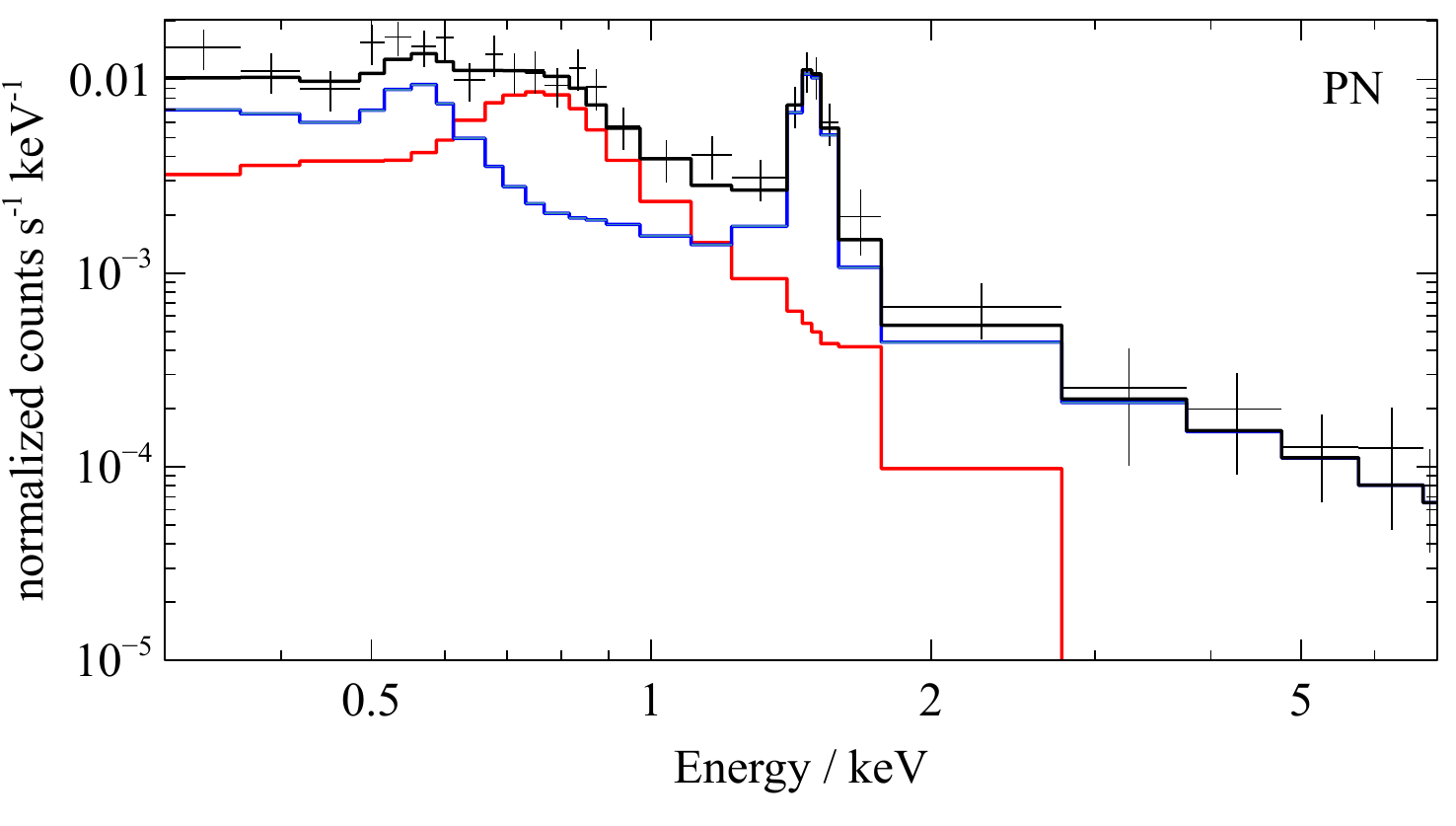}
	\includegraphics[width=\columnwidth]{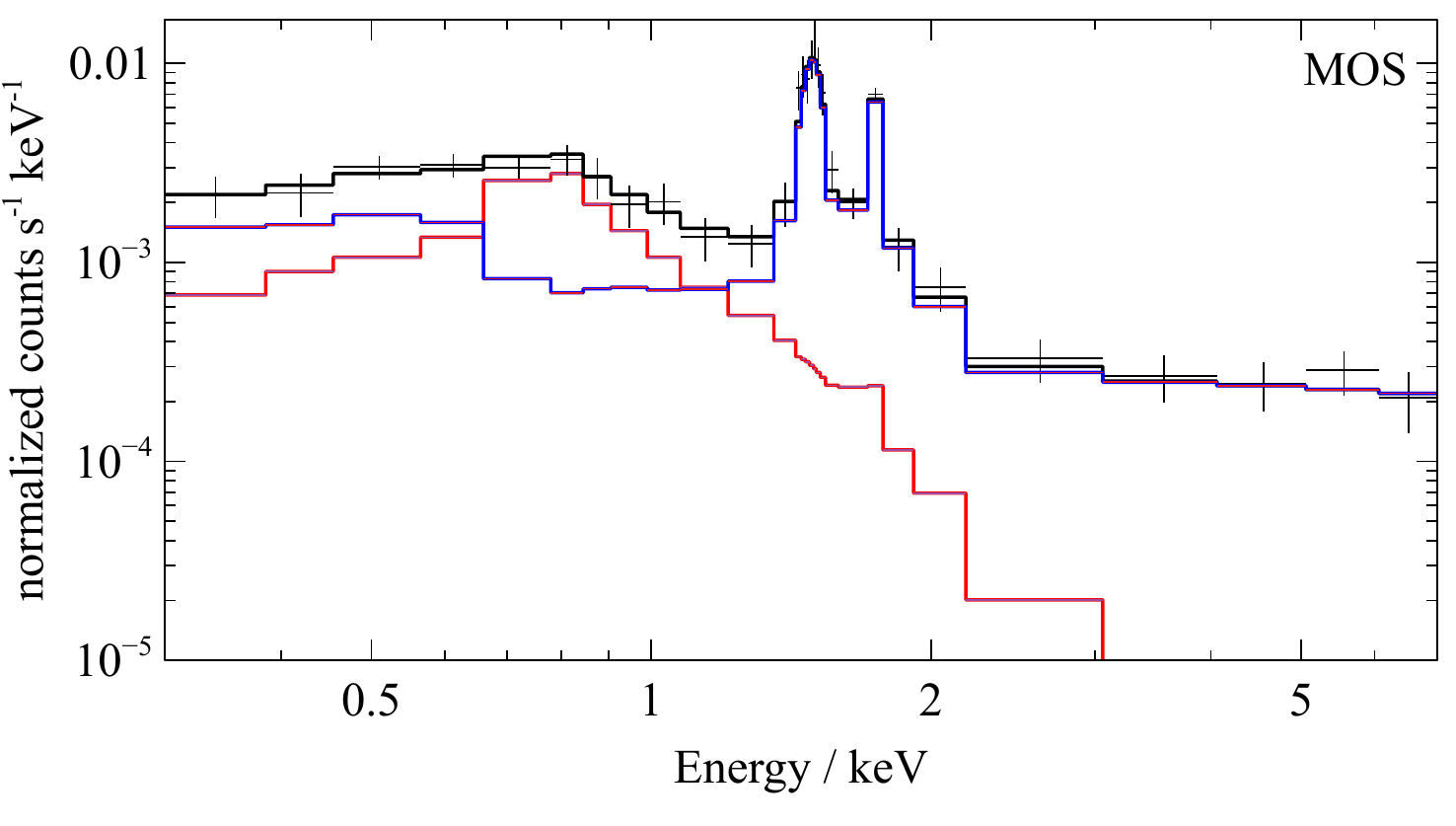}}
	
	\caption{Spectral fits to the PN and MOS data in the 20$-$40 kpc region. Data are shown as black points, the background model is shown as the solid blue line, the thermal component from the galaxy halo is shown as the solid red line, and the total fit (source plus background) is shown as the solid black line. The spectral lines appearing at 1.49 keV and 1.75 keV are the Al K$\alpha$ and Si K$\alpha$ instrumental lines.}
	\label{fig: spectra}
\end{figure}

The emission from the galaxy halo is then modelled as an absorbed APEC component to model the hot gas, and an absorbed power law with index fixed to 1.56 (as in \citealt{anderson2016deep}) to model X-ray binary emission. When fitting for the hot gas, we fit for the temperature and the metal abundance, using the abundance tables of \citet{Grevesse1998}, as in \citet{bogdan2017}. We are able to fit for the temperature in 3 annuli of width 0.25 arcmin, and the plot of the temperature profile and metal abundance profile is shown in Fig. \ref{fig: temp_profile}. Examples of the spectral fits are shown in Fig. \ref{fig: spectra}.

Our findings are similar to those found for NGC 1961 \citep{anderson2016deep} and NGC 6753 \citep{bogdan2017}, in that we find a low value for the metal abundance of $\sim 0.1 \rm{Z}_{\odot}$ which appears uniform with radius, and a temperature profile that decreases slightly with increasing radius.

%----------------------
%----------------------

%----------------------
%----------------------

\subsection{Halo gas mass}
\label{sec: halo-mass}
\begin{figure}
	\includegraphics[width=\columnwidth]{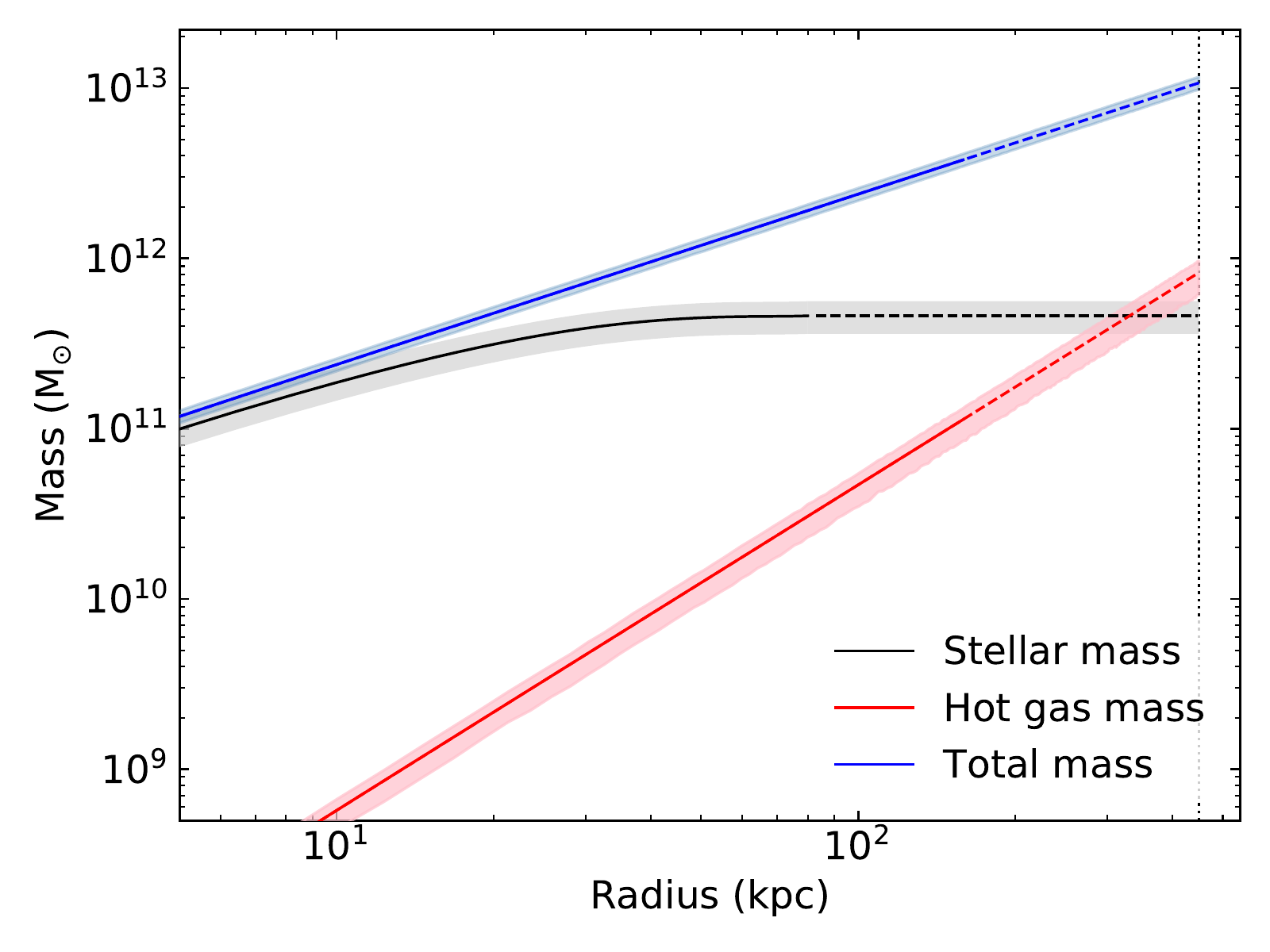}
	\caption{Mass profiles of the spiral galaxy J2345-0449. The black, red, and blue solid lines show, respectively, the distribution of the stellar mass, the hot gas mass, and the total mass. The gas and total masses are measured within a radius of 160 kpc, while the stellar mass is measured within 80 kpc \citep{Walker2015a}. Beyond these radii, mass profiles are extrapolated out to the virial radius (dashed lines). The shadow regions mark the $1\sigma$ error computed using a Monte Carlo technique. The vertical dotted line marks the location of the $r_{200}$ radius.}
	\label{fig: masses}
\end{figure}

To estimate the hot halo mass around the spiral galaxy J2345-0449, we integrated the density profile (equation \ref{eq: density_beta}) over a given volume and multiplied by the mean mass per electron, $\mu_e m_{\rm{p}}$. Using the derived temperature of 0.6 keV and metallicity of 0.1 Z$_{\odot}$, the enclosed gas mass within a radius of 160 kpc, the maximum radius out to which the X-ray emission is detected, is $1.15_{-0.24}^{+0.22} \times 10^{11} \, \rm{M}_{\odot}$. Since the X-ray emission is only detected out to 160 kpc, we must extrapolate the best-fitting density model beyond this radius to estimate the halo mass out to $r_{200}$ (corresponding to 450 kpc). Assuming the gas temperature and metallicity remain constant out to the virial radius, the implied hot halo mass within $r_{200}$ is $8.25_{-1.77}^{+1.62} \times 10^{11} \, \rm{M}_{\odot}$, consistent with a massive hot atmosphere. The best-fitting gas mass profile is shown in Fig. \ref{fig: masses}.

%----------------------
%----------------------

We emphasize that the estimated halo mass within the virial radius has non-negligible systematic and statistical uncertainties, which mainly originate from the fact that the X-ray emission is only detected up to 35 per cent of the virial radius. In addition to the effects of asphericity and the isothermal assumption (see Section \ref{sec: surface_brightness}), the halo mass depends significantly on the solar metallicity. For instance, a higher abundance leads to more emission from the Fe L line due to the degeneracy between the metallicity and the emission measure. For a fixed count rate, this tends to change the gas density, and therefore, the halo mass. Assuming the range of 0.05$-$0.2 Z$_\odot$ for the metallicity, the corresponding possible range for the gas mass within $r_{200}$ is (7.20$-$9.35) $\times 10^{11} \, \rm{M}_{\odot}$. 

\subsection{Total mass}
\label{sec: tot-mass}
The distribution of the galaxy's total mass, which is mainly made up of dark matter, can be computed from the modelled gas structure. Its distribution relates to the density and temperature profiles of the gas. We assume that the gas is in hydrostatic equilibrium in the galaxy’s gravitational potential with no bulk flow of matter. Under the assumptions of isothermal gas and spherical symmetry, the total mass of a galaxy within a radius of $r$ takes the form:
%----------------------
\begin{equation}
    M(r)=\frac{3kT\beta}{G\mu m_{\rm{p}}}\frac{r^3}{r_{0}^2+r^2},
    \label{eq: tot_mass}
\end{equation}
%----------------------
where $G$ is the Newtonian gravitational constant, and $\mu m_{\rm{p}}$ is the mean molecular weight of the gas. Equation \ref{eq: tot_mass} depends on the gas temperature and the halo shape parameters ($\beta$, $r_0$), but it is independent of the value of the central gas density $n_0$.

Within 160 kpc, the derived total mass of the galaxy J2345-0449 is $3.69_{-0.33}^{+0.32} \times 10^{12} \, \rm{M}_{\odot}$, using the estimated gas temperature of 0.6 keV. The best-fitting total mass profile is shown in Fig. \ref{fig: masses}.  When this profile is extrapolated out to $r_{200}$, the derived total mass is $1.07_{-0.09}^{+0.09} \times 10^{13} \, \rm{M}_{\odot}$. Similar to the gas mass measurements, the estimated total mass within $r_{200}$ is also affected by systematic and statistical uncertainties, originating mainly from the unexplored properties of the hot halo beyond a radius of 160 kpc.

\subsection{Magnetic field strength}
\begin{figure}
\centering
	\includegraphics[width=0.6\columnwidth]{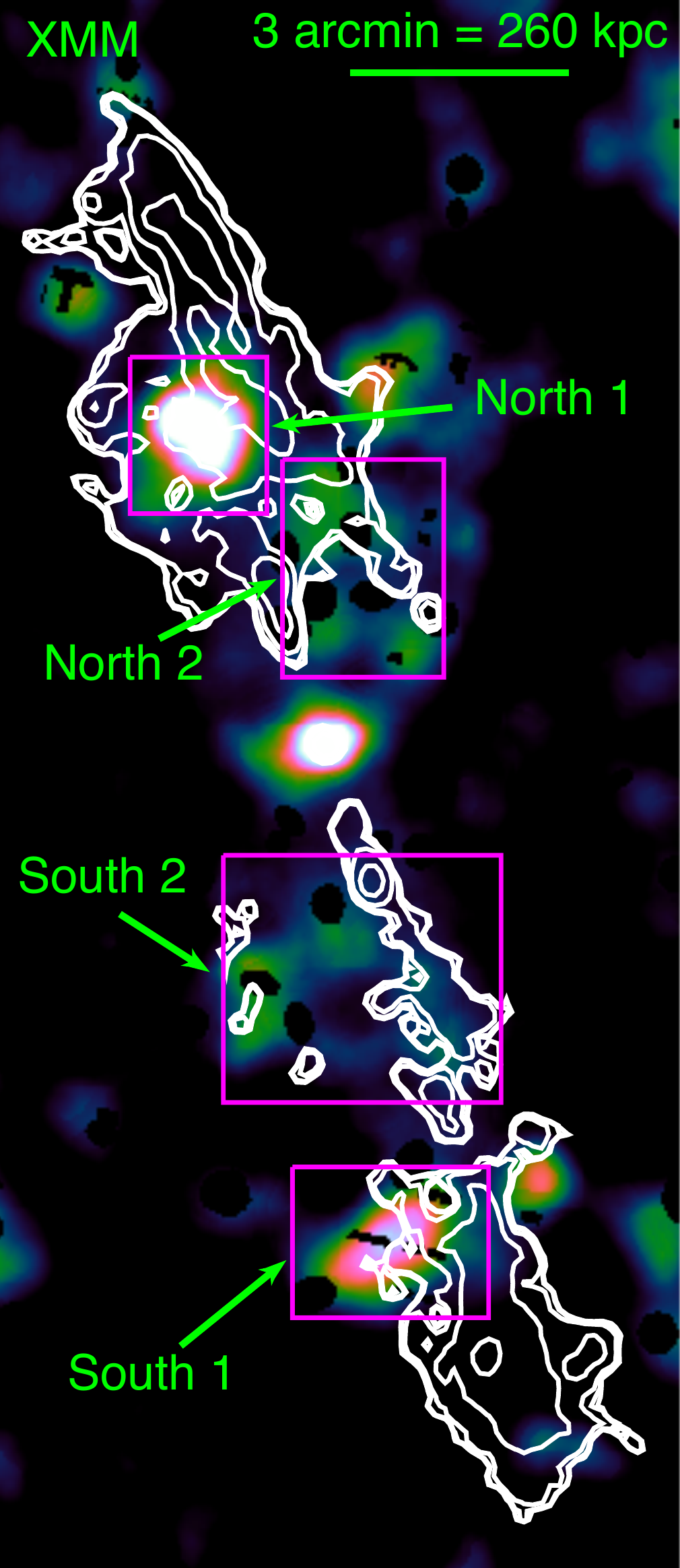}
	\caption{\textit{XMM$-$Newton} image of the spiral galaxy J2345-0449 shows the locations of the regions in which the X-ray and radio fluxes were measured. The GMRT contours are plotted on the top of the X-ray image.}
	\label{fig: lobes}
\end{figure}

We estimated the strength of the magnetic field in the regions where diffuse X-ray emission is associated with the northern and southern radio lobes. Everywhere else where radio emission is seen but not X-ray emission, the magnetic field strength is expected to be relatively high. To estimate the magnetic field strength, we measured the radio flux in four regions using GMRT at 610 MHz, as indicated by magenta boxes in Fig. \ref{fig: lobes}, and subtracted from the background flux, which is measured in a region far away from the galaxy and jets. For X-ray, we extracted the X-ray counts from PN and MOS event data in the same regions, using the XMM-SAS task \textit{xmmselect}. X-ray counts were also extracted from the local background away from the galaxy and jets. For each background-subtracted spectral channel, the spectrum is re-binned so that every bin contains at least 25 counts and not to oversample the intrinsic energy resolution by a factor greater than 3. We fitted the X-ray spectrum with a power-law model, with the normalization (Norm) and spectral index ($\Gamma$) left free to vary. The absorbing column density is fixed to the LAB survey value of $3 \times 10^{20}$ cm$^{-2}$ \citep{kalberla2005leiden}. Values of the best-fitting parameters and the associated $\chi^2$ per degree of freedom for the power-law model are shown in Table \ref{table: best_fitting_parameters}. For each region, the photon index $\Gamma$ estimated from the X-ray spectra is statistically consistent with the radio spectral index ($\Gamma_{\rm{R}}$) value of 2 derived for the northern and southern lobes in the 325$-$1400 MHz range \citep{Bagchi2014}, suggesting that the X-ray emission is attributed to the IC scattering of the CMB photons by a population of the radio-synchrotron-emitting electrons filling the lobes. 

%----------------------
%----------------------

%----------------------------
\begin{table}

    \centering
    \caption{Values of the best-fitting parameters for the power-law model}
    \begin{tabular}{cccc}
    \hline
    Region   &  $\Gamma$ & Norm & $\chi^2$/dof \\
             &    & (Photons keV$^{-1}$ cm$^{-2}$ s$^{-1}$)   &   \\  
    
        \hline
North 1  & $2.10 \pm 0.15$  & $(5.65 \pm 0.39) \times 10^{-6}$ & 102.7/98   \\
North 2  & $2.13 \pm 0.18$  & $(6.62 \pm 0.47) \times 10^{-6}$ & 152.4/124   \\
South 1  & $2.31 \pm 0.36$  & $(8.37 \pm 0.78) \times 10^{-6}$ & 116.2/93  \\
South 2  & $2.25 \pm 0.26$  & $(11.29 \pm 0.68) \times 10^{-6}$ & 179.5/185   \\
  \hline
    \end{tabular}
    \label{table: best_fitting_parameters}
\end{table}
%----------------------------

Using the measurements of the flux density estimated from X-ray and radio data, the magnetic field strength can then be estimated as \citep{harris1979prospects}
%----------------------
\begin{equation}
    B^{\alpha+1}=\frac{(5.05 \times 10^4)^{\alpha} C(\alpha) G(\alpha) (1+z)^{\alpha+3}}{10^{47}}\frac{S_{\rm{R}}}{S_{\rm{X}}}\bigg(\frac{\nu_{\rm{R}}}{\nu_{\rm{X}}}\bigg)^{\alpha},
    \label{eq: magnetic_field}
\end{equation}
%----------------------
where $\alpha \,(=\Gamma_{\rm{R}}-1)$ is the radio spectral index, $C(\alpha)$ is approximately constant ($\approx 1.15 \times 10^{31}$) over the range $0.5 < \alpha < 2.0$, and $G(\alpha)$ is a correction factor tabulated in \citet{harris1979prospects} ($G \approx 0.5$ at $\alpha = 1$). $S_{\rm{X}}$ and $S_{\rm{R}}$ are the X-ray and radio flux densities at frequencies $\nu_{\rm{X}}$ and $\nu_{\rm{R}}$, respectively.

Table \ref{table: magnetic_field} shows values of the magnetic field in the four regions, where radio and X-ray emissions are nearly coincident. In the same table, we also show the X-ray flux density measured at 1 keV and the radio flux density measured using GMRT at 610 MHz. We note that the average value of the magnetic field strength in the north is around $0.36\, \mu$G, higher than that estimated for the south, where the resulting field strength is $\approx 0.23\, \mu$G.  

%----------------------------
\begin{table}
    \centering
    \caption{Flux densities and magnetic field strengths}
    \begin{tabular}{cccc}
    \hline
    Region   & $S_{\rm{X}}$ & $S_{\rm{R}}$ & $B$ \\
             &  (nJy)  & (Jy)  & ($\mu$G)  \\  
        \hline
North 1  & $3.74 \pm 0.26$ & $0.059 \pm 0.001$ & $0.39 \pm 0.01$ \\
North 2  & $4.32 \pm 0.31$ & $0.048 \pm 0.001$ & $0.33 \pm 0.01$ \\
South 1  & $5.53 \pm 0.52$ & $0.030 \pm 0.001$ & $0.23 \pm 0.01$ \\
South 2  & $7.46 \pm 0.45$ & $0.037 \pm 0.002$ & $0.22 \pm 0.01$  \\
  \hline
    \end{tabular}
    \label{table: magnetic_field}
\end{table}
%----------------------------

\section{Discussion}
\subsection{Baryon mass fraction}
\label{sec: baryon_fraction} 
Measuring the baryon mass fraction of galaxies has essential implications for our understanding of galaxy formation and evolution. It allows us to examine the effects of AGN and supernova feedback, and assess their role in expelling gas from the galactic potential well or to large radii. There are a number of baryonic mass components that contribute to the overall baryon budget in galaxies. In addition to the stellar mass, it is found that a significant fraction of the baryons resides in the hot gaseous halo surrounding the host galaxy (e.g. \citealt{Anderson2011}; \citealt{Dai2012}; \citealt{Bogdan2013a}, \citeyear{Bogdan2013b}; \citealt{bogdan2017}). Other contributions come from the cold interstellar medium gas, the molecular gas mass, the photoionized gas ($\approx 10^4$ K) of the circumgalactic medium, and the warm-hot gas ($10^5{-}10^7$ K) of the circumgalactic medium.    

It is expected that the baryon mass fraction in galaxies and clusters to reflect the universal baryon fraction \citep[e.g.][]{white1993baryon}. Yet, the measured baryon mass fraction in galaxies and their halos lies significantly below the universal value \citep[e.g.][]{Anderson2011,bogdan2017}. Cosmological simulations of structure formation \citep[e.g.][]{Sommer-Larsen2006,Fukugita2006} predict that the bulk of the missing baryons resides around the host galaxy, inside of the virial radius, in the form of hot gaseous halos. However, due to its very-low gas density, current X-ray observations are limited to a radius less than 20 per cent of the virial radius, implying that the majority of the halo volume remained undetected. For instance, only 17 per cent of the virial radius is detected in the spiral galaxy NGC 266 (\citealt{Bogdan2013b}), 16 per cent is detected in NGC 1961 \citep{anderson2016deep}, and 11 per cent is detected in NGC 6753 \citep{bogdan2017}.    
Here, we robustly detected diffuse X-ray emission from the hot gaseous halo around the galaxy out to a radius of 160 kpc, corresponding roughly to 35 per cent of the virial radius. To compute the galaxy's baryon content, we only consider the stellar mass and the hot halo gas mass. Contributions from other baryonic components such as the molecular gas mass, the photoionized gas, and the warm-hot gas of the circumgalactic medium are not considered, as they comprise a small fraction of the overall baryon budget in galaxies \citep[e.g.][]{werk2014cos,li2018baryon}. Using IRAM-30m observations, \citet{dabhade2020sagan} reported molecular gas mass of twelve giant radio galaxies, including J2345-0449, and found that this galaxy has a molecular mass of $1.62 \times 10^{10} \, \rm{M}_{\odot}$, much smaller than the stellar mass of the galaxy.   

\citet{Walker2015a} estimated the stellar mass of J2345-0449 using the total K-band absolute magnitude of $-26.15$ determined from 2MASS data. Using a mass-to-light ratio of 0.78 with range $0.60{-}0.95$ \citep{Bell2003,Dai2012}, \citet{Walker2015a} found that the stellar mass of J2345-0449 is $4.6_{-1.0}^{+1.0} \times 10^{11} \, \rm{M}_{\odot}$. Combining this mass with the hot gas mass of $8.25_{-1.77}^{+1.62} \times 10^{11} \, \rm{M}_{\odot}$ estimated within the virial radius (Section \ref{sec: halo-mass}), we find that the total baryon mass is $M_{\rm{b,tot}}=1.29_{-0.39}^{+0.38} \times 10^{12} \, \rm{M}_{\odot}$. Defining the baryon mass fraction as $f_{\rm{b}}=M_{\rm{b,tot}}/(M_{\rm{DM}}+M_{\rm{b,tot}})$, where $M_{\rm{DM}}+M_{\rm{b,tot}}$ is the total mass (Section \ref{sec: tot-mass}), we find that $f_{\rm{b}}= 0.121_{-0.043}^{+0.043}$. The uncertainty associated with the reported baryon mass fraction measurement takes account of the statistical and systematic uncertainties due to the effects of the ellipsoidal shape of the galaxy, and the isothermal and constant metal abundance assumptions. Although the associated uncertainty is large, it is statistically in agreement with the universal baryon fraction of 0.156 $\pm$ 0.003 reported by \citet{ade2016planck}. 

In addition to measuring the baryon mass fraction within the virial radius, we measured the baryon mass fraction within a radius of 160 kpc, the maximum radius out to which the X-ray emission is detected. Within this radius, the total baryon mass is $M_{\rm{b,tot}}=5.75_{-1.73}^{+1.66} \times 10^{11} \, \rm{M}_{\odot}$. Using the estimated total mass of $3.69_{-0.33}^{+0.32} \times 10^{12} \, \rm{M}_{\odot}$, hence, the baryon mass fraction is $f_{\rm{b}}= 0.156_{-0.053}^{+0.051}$, agreeing well with the universal baryon fraction. Here also, we take account of the statistical and systematic uncertainties associated with the baryon mass fraction measurements.

\begin{figure}
	\includegraphics[width=\columnwidth]{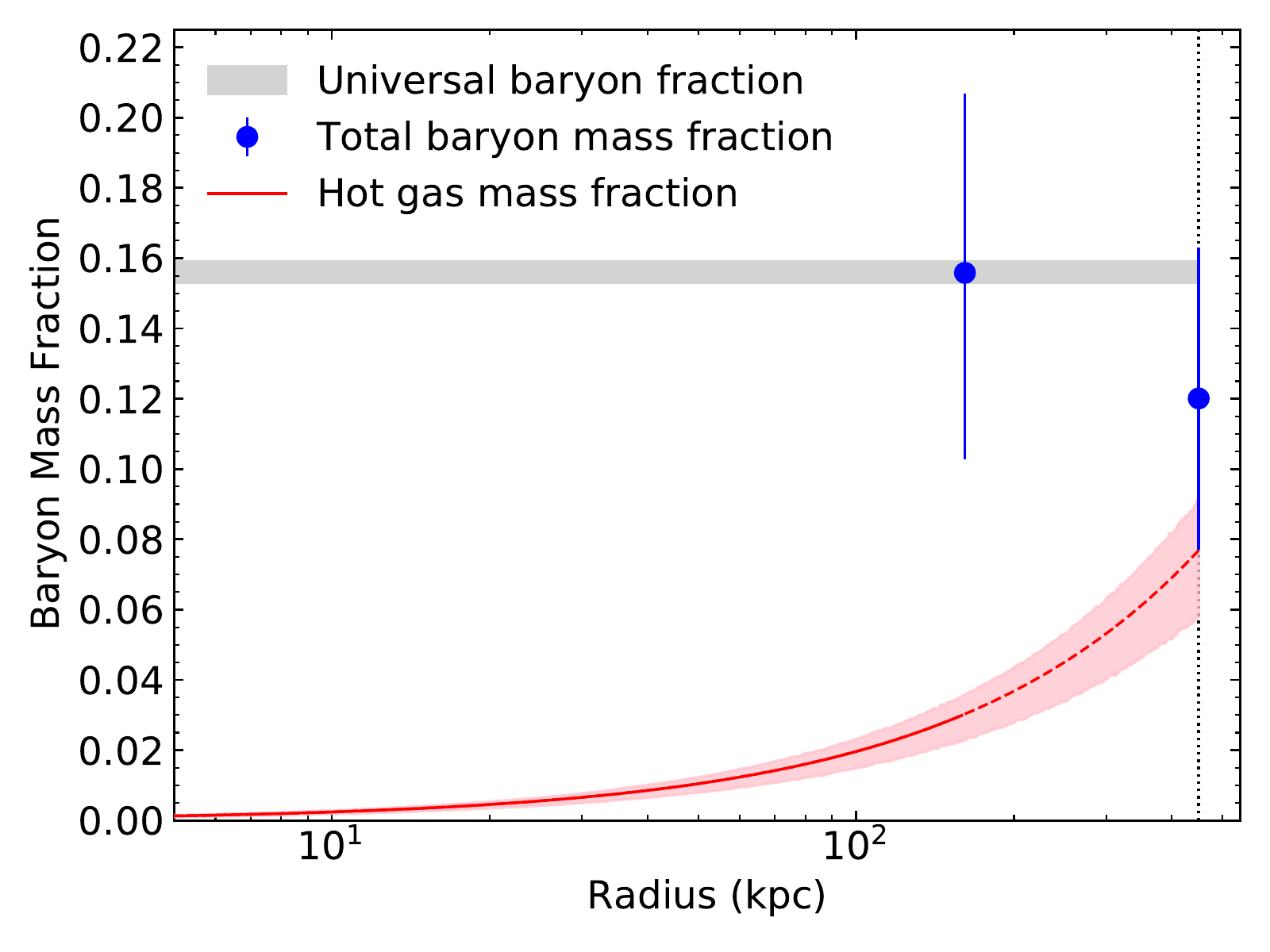}
	\caption{Radial profile of the baryon mass fraction of the spiral galaxy J2345-0449. The red solid line shows the baryon mass fraction profile, only including the hot gas mass. The baryon mass fraction is measured within a radius of 160, and beyond which we extrapolate the profile to the virial radius (dashed line). The shadow pink region marks the $1\sigma$ error computed using a Monte Carlo technique. The blue points mark the estimated values of the baryon mass fraction of J2345-0449 within 160 kpc and 450 kpc, including the stellar mass component estimated by \citet{Walker2015a}. The baryon mass fraction measurements reported at 160 kpc and 450 kpc take into account the systematic and statistical uncertainties due to the effects of the ellipsoidal shape of the galaxy, and the isothermal and constant metal abundance assumptions. The gray region shows the universal baryon fraction \citep{ade2016planck}. The vertical dotted line marks the location of the $r_{200}$ radius.}
	\label{fig: baryon_fraction}
\end{figure}

In Fig. \ref{fig: baryon_fraction}, we show the estimated values of the total baryon mass fraction of the spiral galaxy J2345-0449 within 160 kpc and 450 kpc (the blue points). In the same figure, we also show the radial profile of the hot gas mass fraction of J2345-0449 (the red curve). Within the virial radius (450 kpc), the hot halo mass makes up roughly 65 per cent of the total baryon mass content of the galaxy J2345-0449. The detailed mass budget of J2345-0449 within 160 kpc and 450 kpc is summarized in Table \ref{table: mass_budget}. 

\begin{figure*}
\hbox{	\includegraphics[width=\columnwidth]{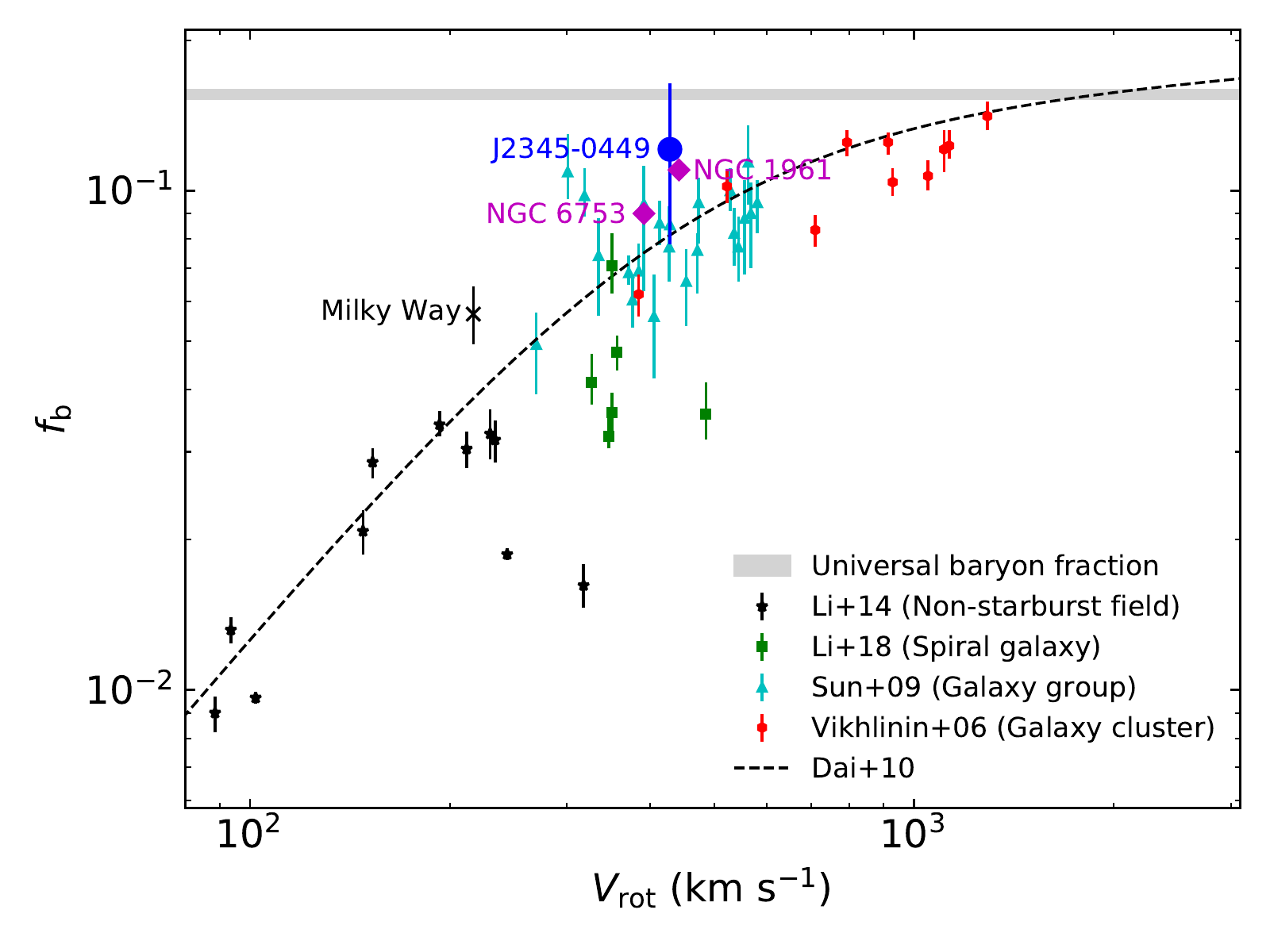}
	\includegraphics[width=\columnwidth]{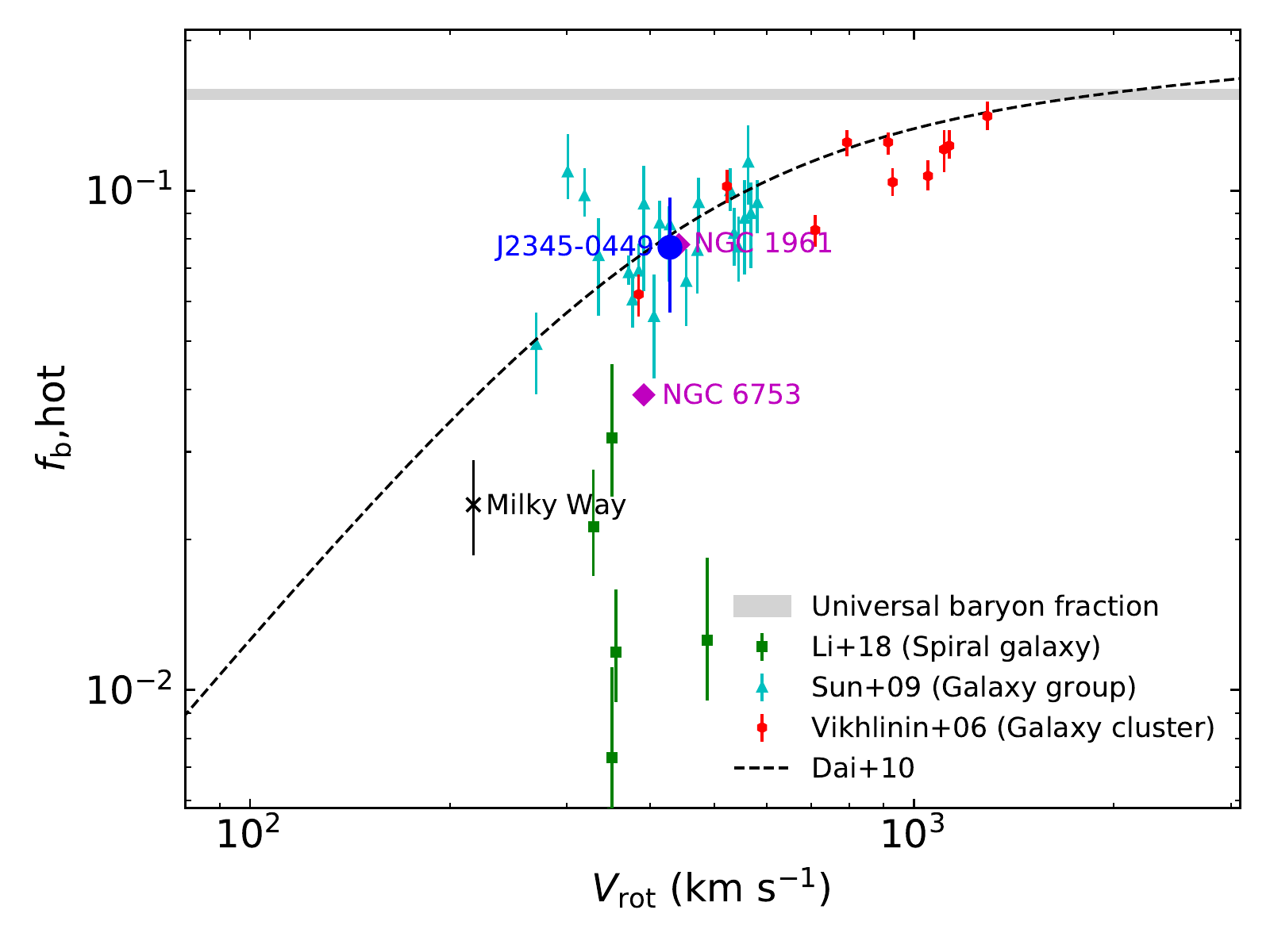}}
	
	\caption{\textit{Left:} Baryon mass fraction vs. rotation velocity. \textit{Right:} Hot gas mass fraction vs. rotation velocity. The J2345-0449 measurements are compared to those reported for a sample of spiral galaxies \citep{li2018baryon}, spiral galaxies NGC 1961 and NGC 6753 (\citealt{Bogdan2013a}), non-starburst field spiral galaxies \citep{li2014chandra}, the Milky Way \citep{miller2015constraining}, galaxy groups \citep{sun2009chandra}, and galaxy clusters \citep{Vikhlinin2006}. The dashed curve is a broken power-law model from \citet{dai2010baryon}. The gray shadow area represents the \textit{Planck} universal baryon fraction \citep{ade2016planck}. The uncertainties associated with the J2345-0449 measurements take account of the systematic and statistical uncertainties discussed in the main text.}
	\label{fig: comparison}
\end{figure*}

We compared the baryon and hot gas mass fraction measurements of J2345-0449 within $r_{200}$ to those reported for a sample of spiral galaxies \citep{li2018baryon}, spiral galaxies NGC 1961 and NGC 6753 (\citealt{Bogdan2013a}), non-starburst field spiral galaxies \citep{li2014chandra}, the Milky Way \citep{miller2015constraining}, galaxy groups \citep{sun2009chandra}, and galaxy clusters \citep{Vikhlinin2006} (see Fig. \ref{fig: comparison}). Results suggest that our estimated baryon and hot gas mass fraction measurements are consistent with those derived for massive spiral galaxies NGC 1961 and NGC 6753 (\citealt{Bogdan2013a}). Our measurements are also consistent with the broken power-law relation between the baryon mass fraction and the rotation velocity found by \citet{dai2010baryon} for galaxy groups and clusters of galaxies (the dashed curve in Fig. \ref{fig: comparison}).

Furthermore, we compared the X-ray luminosity of this spiral galaxy with other massive spiral galaxies. The measured luminosity of the hot halo for J2345-0449 within a radius of 50 kpc in the 0.5$-$2.0 keV energy band is $(4.0 \pm 0.5) \times 10^{41}$ erg s$^{-1}$, significantly higher than the luminosity of $(3.1 \pm 0.5) \times 10^{40}$ erg s$^{-1}$ estimated for NGC 6753 \citep{Bogdan2013a}, and the luminosity of $2.9 \times 10^{40}$ erg s$^{-1}$ estimated for UGC 12591 \citep{Dai2012}. The luminosity of this object is even higher than the luminosity of $(8.9 \pm 1.2) \times 10^{40}$ erg s$^{-1}$ estimated for the luminous giant spiral galaxy NGC 1961 within 50 kpc in the energy band 0.5$-$2.0 keV \citep{anderson2016deep}.

There is a wide range of predictions as to where to find the missing baryons (for a review see \citealt{bregman2018extended}). For the low-feedback cases, one prediction for the missing baryons is that they lie within $r_{200}$. However, most current models do not support that conclusion. Higher feedback cases predict that a significant fraction of baryons lies beyond $r_{200}$. Also, the stacked Sunyaev Zel’dovich signal toward galaxies indicates that the bulk of the missing baryons lies beyond $r_{200}$ \citep{ade2013planck_galaxy}. Our measured baryon mass fraction is consistent with all baryons falling within $r_{200}$, or with only half of the baryons falling within $r_{200}$.

%----------------------
%----------------------

%----------------------------
\begin{table*}
\begin{minipage}{130mm}
    \centering
    \caption{Mass budget of J2345-0449}
    \begin{tabular}{cccccc}
    \hline
    Radius   &  $M_{\star}$ & $M_{\rm{gas}}$ & $M_{\rm{b,tot}}$ & $M_{\rm{DM}}+M_{\rm{b,tot}}$ & $f_{\rm{b}}$ \\
    (kpc)  &  ($\rm{M_{\odot}}$)  &   ($\rm{M_{\odot}}$) &  ($\rm{M_{\odot}}$) &  ($\rm{M_{\odot}}$)  &  \\
        \hline
160  & $4.6_{-1.0}^{+1.0} \times 10^{11}$ & $1.15_{-0.24}^{+0.22} \times 10^{11}$ & $5.75_{-1.73}^{+1.66} \times 10^{11}$ & $3.69_{-0.33}^{+0.32} \times 10^{12}$  & $0.156_{-0.053}^{+0.051}$  \\
450  & $4.6_{-1.0}^{+1.0} \times 10^{11}$ & $8.25_{-1.77}^{+1.62} \times 10^{11}$ & $1.29_{-0.39}^{+0.38} \times 10^{12}$ & $1.07_{-0.09}^{+0.09} \times 10^{13}$  & $0.121_{-0.043}^{+0.043}$  \\
  \hline
    \end{tabular}
    %\vspace{1ex}
    
  {\raggedright \textbf{Notes.} From left to right, columns are as follows: (1) Radius within which the masses are measured. (2) Estimated stellar mass within 80 kpc \citep[from][]{Walker2015a}. (3) Estimated hot gas mass. (4) Total baryon mass, which is the sum of the stellar mass and the hot gas mass. (5) Estimated total mass, which is computed using equation (\ref{eq: tot_mass}). (6) Estimated baryon mass fraction, defined as $f_{\rm{b}}=M_{\rm{b,tot}}/(M_{\rm{DM}}+M_{\rm{b,tot}})$. The uncertainties associated with the baryon mass fraction take account of the systematic and statistical uncertainties discussed in the main text. \par}
    \label{table: mass_budget}
    \end{minipage}
\end{table*}
%----------------------------  

%----------------------
%----------------------

\subsection{Energetics in the lobes}
The electron and magnetic energies stored in the radio lobes are sourced by jets originating in the nucleus of the radio galaxy. This makes the lobes to be a valuable probe to trace the final stage of the evolution of radio jets. Typically, the energy content of the radio lobes is measured via the diffuse radio synchrotron emission and the X-ray IC scattering of CMB photons by a population of the relativistic electrons injected into the radio lobes. In many systems, the lobes have been claimed to be dominated by electrons, i.e., the magnetic field is lower than the equipartition value \citep[see][]{isobe2002chandra,croston2005x,isobe2006xmm,Stawarz2013giant,isobe2015x}, whereas in others a near-equipartition condition have been claimed between the magnetic field and electron energy densities \citep[see][]{hardcastle2002magnetic,croston2004x,isobe2011suzaku}.

In this paper, we measured the energy densities of the electrons ($U_{\rm{e}}$) and magnetic field ($U_{\rm{B}}$) in the regions shown as magenta boxes in Fig. \ref{fig: lobes}, using analytical formulas described in \citet{harris1979prospects}. We approximated the shape of these regions to be prolate ellipsoids, and the electron number density was assumed to follow a power-law distribution of electrons: $N(\gamma)=N_0\gamma^{-p}$, where $N_0$ is the amplitude of the electron spectrum, $\gamma$ is the Lorentz factor, and the index $p=2\Gamma_{\rm{R}}-1$. Accordingly, the energy density of the electron population in a radio lobe takes the form:
%----------------------
\begin{equation}
    U_{\rm{e}}=\int_{\gamma_{\rm{min}}}^{\gamma_{\rm{max}}} N(\gamma)\gamma m_{\rm{e}}c^2 {\rm{d}}\gamma,
    \label{eq: electron_energy}
\end{equation}
%----------------------
where $\gamma_{\rm{min}}$ and $\gamma_{\rm{max}}$ are, respectively, the minimum and maximum Lorentz factors, $m_{\rm{e}}$ is the electron mass, and $c$ is the light speed.

The electron energy density, however, is relatively sensitive for different choices of $p$ and $\gamma_{\rm{min}}$. The spectral index $p$ can be constrained observationally, whereas the Lorentz factor $\gamma_{\rm{min}}$ has not been constrained yet. Some studies \citep[e.g.][]{erlund2006extended} found that the minimum electron energy in the radio lobes corresponds to $\gamma_{\rm{min}}=10^3$, while others derived a lower value for $\gamma_{\rm{min}}$ in some giant radio galaxies \citep[e.g.][]{orru2010low}. Here, we estimated the electron energy density by adopting values of $10^3$ and $10^2$ for $\gamma_{\rm{min}}$. 

The results of the estimated electron energy density in these four regions are summarized in Table \ref{table: energetics}. In the same table, we also show the magnetic energy density, which is derived from the magnetic field strength as $U_{\rm{B}}=B^2/2\mu_0$, where $\mu_0$ is the magnetic permeability. From the measured energy densities of the electrons and magnetic field, we have computed the $U_{\rm{e}}/U_{\rm{B}}$ (Table \ref{table: energetics}). Our results indicate that the radio lobes of the galaxy J2345-0449 are electron-dominated by up to a factor of 23 and 197 for $\gamma_{\rm{min}}$ of $10^3$ and $10^2$, respectively.

There are several possible sources of photons that can be IC-scattered up to X-ray energies by a population of the relativistic electrons. These sources include synchrotron photons produced from the radio emission itself in a process known as synchrotron self-Compton \citep[e.g.][]{hardcastle2004origins}, and the CMB radiation \citep[e.g.][]{harris1979prospects}. In the northern and southern lobes of J2345-0449, we found that the energy density of the synchrotron radiation is much smaller than the photon energy density from the CMB radiation ($U_{\rm{CMB}}=5.65 \times 10^{-13}$ erg cm$^{-3}$). This implies that the seed photons are mainly originated from the CMB radiation and the radio synchrotron emission has only a negligible contribution. Furthermore, we found a dominance of the IC radiative lose over that from the synchrotron emission (i.e. $U_{\rm{B}}/U_{\rm{CMB}} \approx 0.01$), agreeing with that found in the lobes of other giant radio galaxies \citep[e.g. 3C 236;][]{isobe2015x}.

%----------------------------
\begin{table*}
    \centering
    \caption{Summary of energetics in the northern and southern lobes}
    \begin{tabular}{cccccc}
    \hline
         &   & \multicolumn{2}{c}{$\gamma_{\rm{e}}=10^3{-}10^5$} & \multicolumn{2}{c}{$\gamma_{\rm{e}}=10^2{-}10^5$}  \\
    \cline{3-6}
    
    Region   & $U_{\rm{B}}$ & $U_{\rm{e}}$ & $U_{\rm{e}}/U_{\rm{B}}$  & $U_{\rm{e}}$ & $U_{\rm{e}}/U_{\rm{B}}$ \\
             &  ($10^{-14}$ erg cm$^{-3}$)  & ($10^{-14}$ erg cm$^{-3}$)  &  & ($10^{-14}$ erg cm$^{-3}$)  &   \\  
        \hline
North 1 &  $0.61 \pm 0.03$ & $8.91 \pm 0.62$ & $14.61 \pm 1.24$ & $70.37 \pm 5.59$ & $115.36 \pm 10.78$ \\
North 2 & $0.43 \pm 0.03$  & $4.13 \pm 0.30$ & $9.61 \pm 0.97$  & $37.23 \pm 2.67$ & $86.58 \pm 8.66$  \\
South 1 & $0.21 \pm 0.02$  & $4.92 \pm 0.46$ & $23.43 \pm 3.13$ & $41.34 \pm 4.17$ & $196.86 \pm 27.31$ \\
South 2 & $0.19 \pm 0.02$  & $3.87 \pm 0.23$ & $20.37 \pm 2.46$ & $34.87 \pm 2.10$ & $183.53 \pm 22.26$ \\
  \hline
    \end{tabular}
    \label{table: energetics}
\end{table*}
%----------------------------

\section{Conclusions}
We have presented a deep \textit{XMM$-$Newton} observation of the extremely massive, rapidly  rotating, relativistic-jet-launching spiral galaxy J2345-0449. We have been able to detect diffuse X-ray emission from the hot gaseous halo surrounding the galaxy out to about 160 kpc, out to about 35 per cent of $r_{200}$. Beyond this radius, we have extrapolated our measurements to the virial radius. Furthermore, we have also detected X-ray emission associated with the northern and southern lobes, possibly attributed to IC scattering of CMB photons by a population of the radio-synchrotron-emitting electrons.

Fitting X-ray emission with the standard isothermal $\beta$ model, we have found that the enclosed gas mass within 160 kpc, the maximum radius out to which the X-ray emission is detected, is $1.15_{-0.24}^{+0.22} \times 10^{11} \, \rm{M}_{\odot}$. When we have extrapolated the gas mass profile out to the virial radius, the enclosed gas mass is $8.25_{-1.77}^{+1.62} \times 10^{11} \, \rm{M}_{\odot}$, corresponding roughly to 65 per cent of the total mass in baryons within the virial radius of the galaxy. This is consistent with the predictions of cosmological simulations \citep[][]{Sommer-Larsen2006,Fukugita2006}. With including the stellar mass component and accounting for the statistical and systematic uncertainties, we have found that the baryon mass fraction is $0.121_{-0.043}^{+0.043}$ within the virial radius, statistically in agreement with the universal baryon fraction and massive spiral galaxies NGC 1961 and NGC 6753. We have also measured the energy densities of the electrons and magnetic field for the detected X-ray emission regions associated with the northern and southern lobes, and our results indicate that these regions are electron-dominated by a factor of about $10{-}200$, depending on the choice of the lower cut-off energy of the electron population.

\section*{Acknowledgements}
We thank the referee for their helpful report. This work based on observations obtained with \textit{XMM$-$Newton}, an ESA science mission with instruments and contributions directly funded by ESA Member States and NASA. Observations with the NASA/ESA Hubble Space Telescope obtained at the Space Telescope Science Institute, which is operated by the Association of Universities for Research in Astronomy, Inc., under NASA contract NAS 5-26555. These observations are associated with program \#14091. We thank the staff of the GMRT that made these observations possible. GMRT is run by the National Centre for Radio Astrophysics of the Tata Institute of Fundamental Research. LCH was supported by the National Science Foundation of China (11721303, 11991052) and the National Key R\&D Program of China (2016YFA0400702).

%%%%%%%%%%%%%%%%%%%%%%%%%%%%%%%%%%%%%%%%%%%%%%%%%%
\section*{Data Availability}
The \textit{XMM$-$Newton} Science Archive (XSA) stores the archival data used in this paper, from which the data are publicly available for download. The \textit{XMM} data were processed using the \textit{XMM$-$Newton} Science Analysis System (SAS). The software packages \textsc{heasoft} and \textsc{xspec} were used, and these can be downloaded from the High Energy Astrophysics Science Archive Research Centre (HEASARC) software web page. Analysis and figures were produced using \textsc{python} version 3.7.



\bibliographystyle{mnras}
\bibliography{spiral_galaxy}

%\begin{thebibliography}{99}
%\end{thebibliography}

%%%%%%%%%%%%%%%%%%%%%%%%%%%%%%%%%%%%%%%%%%%%%%%%%%

%%%%%%%%%%%%%%%%% APPENDICES %%%%%%%%%%%%%%%%%%%%%

%\appendix


%%%%%%%%%%%%%%%%%%%%%%%%%%%%%%%%%%%%%%%%%%%%%%%%%%

% Don't change these lines
\bsp	% typesetting comment
\label{lastpage}
\end{document}